\documentclass{iaus}
\usepackage{graphicx}

\def\kms{km\,s$^{-1}$}

\def\xmm{{\sc XMM}\emph{-Newton}}

\def\ch{\emph{{\it Chandra}}}

\title[JD 11.~~X-ray stellar population of the LMC] 
{X-ray stellar population of the LMC}

\author[Ya\"el Naz\'e]   
{Ya\"el Naz\'e
  \thanks{Postdoctoral Researcher FRS-FNRS}
}

\affiliation{Institut d'Astrophysique et de G\'eophysique, Universit\'e de Li\`ege, \\All\'ee du 6 Ao\^ut 17, Bat B5C, B4000-Li\`ege, Belgium  \\ email: {\tt naze@astro.ulg.ac.be} }

\pubyear{2008}
\volume{256}  
\pagerange{1--12}
\jname{The Magellanic System: Stars, Gas, and Galaxies}
\editors{Jacco Th. van Loon \& Joana M. Oliveira, eds.}
\begin{document}

\maketitle

\begin{abstract}
In the study of stars, the high energy domain occupies a 
place of choice, since it is the only one able to directly probe the most 
violent phenomena: indeed, young pre-main sequence objects,
hot massive stars, or X-ray binaries are best revealed in X-rays. However, 
previously available X-ray observatories often provided only crude 
information on individual objects in the Magellanic Clouds. The advent of the highly 
efficient X-ray facilities XMM-Newton and Chandra has now dramatically
increased the sensitivity and the spatial resolution available to X-ray 
astronomers, thus enabling a fairly easy determination of the properties 
of individual sources in the LMC.
\end{abstract}

\firstsection 
\section{X-ray surveys of the LMC}
In the range of astronomical tools, X-rays are of particular interest. Indeed, 
the high-energy domain unveils the most energetic phenomena taking place
in our Universe. Such processes are usually difficult to perceive at other 
wavelengths, though they provide important constraints for astrophysics.\\

In this context, the MCs are targets of choice. Their advantages are multiple: 
the known and small distance, together with the small angular size, small inclination, different metallicities, and recent star formation episodes are 
here as crucial as at other wavelengths. In addition, the low obscuration
towards the MCs renders soft X-ray observations much easier while the numerous
available data taken at other wavelengths ensure a correct, global analysis
of the LMC X-ray sources.\\

X-rays associated with the LMC were first detected 40 years ago by a rocket 
experiment (\cite{mar69}): the source appeared extended and soft, with a
total luminosity estimated to 4$\times 10^{38}$\,erg\,s$^{-1}$. It did not 
take long to distinguish a few individual sources in this X-ray emission, nicknamed 
LMC\,X-1 to 6, thanks to the joint effort of the satellites Uhuru, Copenicus, 
OSO-7, and Ariel V (\cite{leo71}, \cite{rap74}, \cite{mar75}, \cite{gri77}). 
In the following decade, the Einstein observatory 
increased the number of known X-ray sources in the direction of the LMC 
to about a hundred (\cite{lon81}, \cite{wan91}). Finally, a sensitive 
survey undertaken by ROSAT provided another ten-fold increase in the total
source number 
(\cite[Haberl \& Pietsch 1999a, Sasaki \etal\ 2000]{hab99a,sas00}).\\

However, only half of the detected X-ray sources truly belonged to the LMC and 
an even smaller fraction appeared to be associated with LMC stars. For example,
of the 758 ROSAT-PSPC sources, only 144 were identified at first 
(\cite[Haberl \& Pietsch 1999a]{hab99a}): 15 as background AGNs or 
galaxies behind the LMC, 57 as 
foreground Galactic stars, 46 as SNRs and SNRs candidates, 17 as X-ray 
binaries (XRBs) and candidates, 9 as Supersoft sources (SSSs) and 
candidates. Using the observed X-ray properties (especially the hardness 
ratios), \cite{hab99a} further proposed additional 
identifications (3 AGNs, 27 foreground stars, 9 SNRs and 3 SSSs) which 
yields a fraction of 20\% of X-ray sources associated with LMC stars. Similar
results were obtained with the ROSAT-HRI (\cite[Sasaki \etal\ 2000]{sas00}): 
397 detections 
among which 138 in common with the PSPC and 115 identified (10 AGNs, 52 
foreground stars, 33 SNRs, 12 XRBs, 5 SSSs, and 3 hard sources which could 
either be AGNs or XRBs). If one considers the variable X-ray sources, the
contamination by non-LMC objects is smaller. For the PSPC survey, the 
proposed identification of the 27 variable X-ray sources is 12 XRBs, 
5 SSSs, 9 foreground stars and 
1 Seyfert galaxy (\cite[Haberl \& Pietsch 1999b]{hab99b}); for the HRI survey, 26 variable 
sources were detected among which 8 XRBs, 4 SSSs, 6 foreground stars, 
2 AGNs and 1 nova (\cite[Sasaki \etal\ 2000]{sas00}): the fraction of LMC stellar 
objects among variable X-ray sources is thus 60\%.\\

The current facilities possess much higher sensitivities and spatial/spectral resolution but unfortunately 
they also have a smaller field-of-view. Its non-zero extension 
penalizes the LMC, especially in comparison with the SMC ($\sim$59 sq. 
deg. vs. $\sim$18). This explains why, up to now, no full survey of the 
LMC has been performed with \xmm\ or \ch. Nevertheless,
smaller fields have been observed and it should be underlined that, though
its coverage is patchy, the 2XMM catalog currently lists 5421 entries 
in the area of the ROSAT surveys! In addition,
\cite{hab03} reported the analysis of one deep \xmm\ 
observation of a northern region of the LMC (on the rim of the supergiant
shell LMC4). While ROSAT had detected 34 sources in this field, \xmm\
data reveal 150 objects (detection limit 6$\times 
10^{32}$\,erg\,s$^{-1}$). In a selection of 20 bright or peculiar sources, 
the majority (10) are AGNs, but there are also 3 foreground stars, 2 SNRs, 
4 HMXBs, and 1 SSSs \footnote{This source was not confirmed by \cite{kah08}.}. In addition, \cite{sht05} analyzed 23 \xmm\ archival observations covering 
3.8 sq. degrees of the LMC. With a detection limit of 3$\times 
10^{33}$\,erg\,s$^{-1}$, they detected 460 sources in the 2--8\,keV band, 
in vast majority AGNs, and focused on 9 good XRBs candidates and 19 possible 
XRBs (see below for more details). 
Finally, a sensitive survey of the LMC in the hard X-ray domain 
(15\,keV--10\,MeV) has been performed with {\sc Integral} (\cite{got06}). 
Only a few sources have been detected: the X-ray binaries LMC\,X-1, 
LMC\,X-4, and PSR B0540--69, as well as two hard sources which might 
correspond to LMC binaries. These encouraging first results in both the soft
and hard X-ray domains enlight the detection potential of the sensitive 
observatories of the current generation. \\

\section{Supersoft X-ray sources}
As their names indicate, SSSs display very soft spectra: they can be 
fitted with thermal models with kT of only 20--80\,eV. These often 
bright (10$^{36-38}$\,erg\,s$^{-1}$) objects became a category 
of their own after the discovery of several such sources in the LMC - indeed, 
their identification is most difficult in the Galatic plane, where high 
absorption prevents their detection. Including recent \xmm\ detections,
18 SSSs are now identified in the LMC and several other 
examples are also known in $>$10 galaxies (\cite[Kahabka \etal\ 2008]{kah08}). \\

It is generally believed that the X-ray emission of SSSs corresponds to 
steady nuclear burning on the surface of a white dwarf accreting matter
from its companion. SSSs would therefore trace rather old stellar populations,
and they are accordingly found along the rim of the LMC optical 
bar (Haberl \& Pietsch 1999a,b). Other possible origins for SSSs have been
proposed and the LMC SSSs confirmed this fact: of the 18 known SSSs, 5 (+1?) 
reside in close binaries (as expected from the above picture), but 5 others 
correspond respectively to a post-nova object, a WD+Be binary, a symbiotic 
star, a planetary nebula, a transition (WD--central star of PN) object - 7 remain unidentified 
transients (\cite[Kahabka \etal\ 2008]{kah08}). \\ 

The X-ray spectrum of SSSs often presents `photospheric' characteristics, 
i.e. it consists of numerous absorption lines from highly ionized metals 
(generally Si, S, Ar, Ca, Fe) superimposed on a blackbody emission. 
One important contribution from \xmm\ and \ch\ is the availability of 
high-resolution spectroscopy. Combining high-res spectra and (preferentially NLTE) 
atmosphere modelling, it is possible to derive precise stellar parameters, 
notably the WD mass. Such an analysis was undertaken for CAL83 by 
\cite{lan05}, who found M$_{WD}$=1.3$\pm$0.3\,M$_{\odot}$ and T=0.55$\pm$0.03\,MK.
Many absorption lines were still blended, preventing any detailed chemical 
analysis. Looking closer at the lines, there seems to be no evidence for
emission components, large line shifts or asymmetries (which would indicate 
fast outflows), or large line widths (which would be associated with a fast 
rotation). A similar study of the SSS associated with nova LMC 1995 led to 
M$_{WD}$=0.91\,M$_{\odot}$ and 
T=0.40--0.47\,MK (\cite{ori03}) and showed that the abundance of carbon 
is not enhanced, suggesting the X-ray emitting matter to be accreted material
from the companion. \\

Some SSSs present however very different spectral characteristics. At high
resolution, the X-ray spectrum of CAL87 appears composed exclusively of 
numerous $emission$ lines (mostly from O, N, and Fe). These lines
which come from recombination and resonant scattering are clearly 
redshifted, with a double-peaked structure observable in 
the brightest features (\cite{gre04}, \cite{ori04}). 
In addition, the broad X-ray eclipse suggests the X-ray source to be 
extended, about 1.5\,R$_{\odot}$ in size (to be compared to the orbital
separation of 3.5\,R$_{\odot}$). The X-rays thus originate in a fast 
(2000\,\kms) and non-spherical outflow, e.g. an accretion disk corona.\\

The brightest SSS is RX\,J0513.9--6951. Its spectrum is a mixture of 
absorption and emission lines superimposed on a continuum, 
indicating the superposition of emissions from a hot atmosphere and from
an optically-thin corona (\cite{mcg05}, \cite{bur07}). 
The absorption lines exhibit several blueshifted velocity components, a 
typical signature of an outflow, which vary with time (the deepest 
absorptions were observed when the X-ray flux was lowest). High abundances
are observed for N, S, and Ar; they may imply that the outflowing
material was affected by nova nucleosynthesis (\cite{mcg05}).
Two temperature components may in fact be present and would be linked to
a fast rotation of the WD (cool equatorial regions + hot polar caps, 
\cite{bur07}). \\

SSSs are also variable X-ray sources: they exhibit a variety of flux changes 
occuring on a variety of timescales. For example, CAL83 displays recurrent 
X-ray low states and short-term variations of smaller amplitude 
(\cite[Lanz \etal\ 2005]{lan05}). The presence of 38m oscillations at some epochs was also 
claimed for this object by \cite{sch06} 
who attributed them to non-radial pulsations. Another SSS,
RX\,J0513.9--6951, displays every 100--200d optically faint/X-ray bright 
episodes of duration 20--40d. The exact 
recurrence timescale varies, which is probably related to variations
in the accretion rate by a factor of 5 (\cite{bur08}). The high-res X-ray
data were acquired during one of the low optical state but the last \xmm\ 
observation samples the beginning of the recovering to `normal' 
optical/UV intensity (\cite{mcg05}, \cite{bur07}).
As the recovering progresses, the X-ray luminosity decreases, as well as
the temperature of the blackbody: it thus seems that the peak emission
slowly shifts towards longer wavelengths (\cite{mcg05}). 
At the same time, the radius of the blackbody emitter becomes larger. This is 
consistent with the global picture of the WD contraction model: the accretion 
rate is high during the low optical state; when it drops, the WD contracts 
and the emission becomes more energetic; the enhanced X-ray emission
then influences the WD environment and provokes a new increase in the 
accretion rate, thereby inflating again the WD (\cite{mcg05}). 
Another interpretation implies changes in the accretion disk's size and
the wind outflow, but observations seem to contradict the predictions
from that particular model (\cite{bur07}).

\section{X-ray binaries}

\subsection{HMXBs}
In high-mass X-ray binaries, the primary is a compact object 
(neutron star, NS, or black hole, BH) while
the secondary can either be a Be star or a supergiant star. In the first 
case, the binary is generally eccentric and accretion onto the compact 
companion occurs only when the latter crosses the dense equatorial 
regions of the Be star, producing recurrent X-ray outbursts. Most HMXBs 
of the LMC are of this type. In the second case, when the secondary is a 
supergiant, the accretion occurs either through wind capture or Roche-lobe 
overflow. In the LMC, there are one or two candidates for wind accretion 
and two or three candidates for RLOF (see e.g. \cite{neg02}).\\

As they contain massive secondaries, HMXBs reveal young stellar populations.
However, they are not instantaneous tracers of star formation, since 
one needs to wait until the compact object forms: while the number of HMXBs 
certainly decreases for populations older than 20Myr, no HMXB is expected 
when the stellar population is younger than 3Myr. This explains why 30 
Doradus, one of the most active star-forming regions in the LMC, contains
few HMXBs while the $\sim$10\,Myr supergiant shell LMC4 harbors many of them
(Haberl \& Pietsch 1999a,b, \cite[Shtykovskiy \& Gilfanov 2005]{sht05}).\\

While only 10 HMXBs were known in the MCs before 1995, their number now
exceeds 100. In the LMC, 36 cases are listed by \cite{liu05}. 
\cite{sht05} studied the cumulative distribution of 
X-ray luminosities (the X-ray luminosity function) of HMXBs and candidates 
found in 23 \xmm\ fields. They reported a possible flattening of the
luminosity function at high luminosities. If confirmed, this would indicate
a deficit of low-luminosity sources, probably linked to the `propeller effect'
(i.e. inhibition of accretion, and therefore decrease of the X-ray luminosity,
by fast-rotating NS with low accretion rates).\\

Among the HMXBs of the LMC, two could contain black holes: LMC\,X-1 
and LMC X-3 (m$_{BH}\sim4$\,M$_{\odot}$, see e.g. \cite{yao05}). Such systems 
display two spectral states: a high/soft state and a low/hard state.
In the former case, the X-ray luminosity is high while the X-ray spectrum 
is soft and the X-ray emission is believed to be mostly thermal emission 
from the accretion disk.
In the latter case, the situation is opposite and the X-ray emission
is attributed to comptonization of the soft photons from the accretion 
disk by hot electrons in the surrounding corona. In most cases, the spectrum 
is thus fitted by the combination of a multi-temperature blackbody model 
(where the temperature decreases while the radius increases) and a power-law. 
This yields rather good results, especially when the photon statistics is poor, 
but is clearly oversimplistic (presence of an intense gravitational 
field in the inner 
regions, interrelations between the comptonized emission and the seed 
blackbody spectrum,...) and can be improved in many ways (\cite{yao05} 
and references therein).\\

LMC\,X-1 comprises an O8III-V star in a 4.2d orbit around a BH (\cite{neg02}). 
Up to now, it has only been observed in the high/soft state
but variations of its X-ray emission were detected, even for the
hardest X-rays (detection in the 20--40\,keV range in 2003 but not in 2004, 
\cite{got06}). Only a disk blackbody was requested 
to fit its spectrum and no discrete features were found in the high-resolution 
X-ray spectrum, though one would have expected the Ne\,K edge to be detected
(\cite{cui02}). \\

LMC\,X-3 harbours a B star of uncertain type (B2.5V, \cite{neg02}, 
or B5IV, \cite{wu01}) in a 1.7d orbit around a BH. This system is often 
- but not always - seen in the high/soft state and a combination of a
multicolour blackbody and a power law is needed to describe its X-ray 
spectrum (\cite{cui02}). Peaks in the power-law flux are related to 
drops in the blackbody flux: accreting matter thus appears diverted from 
the accretion disk to feed the surrounding hot corona (\cite{smi07}). 
The transfer of material occurs most probably
through Roche lobe overflow since there is no evidence for the presence of
a wind (no emission lines, no absorption edges, and low absorbing column for 
neutral/ionized matter compatible with Galactic values, \cite{wu01}, 
\cite{pag03}). Turning to the X-ray lightcurve of LMC\,X-3, there
appear to be no significant variations on very short timescales; 
however, modulations with 
the orbital period and with twice this period were detected (\cite{boy01}).
The former, which is particularly obvious when the X-ray flux is lower, 
could be attributed to the presence of a hot spot (e.g. where the gas 
stream impacts the accretion disk). On the other hand, the abrupt flux 
decrease seen each 2$\times$P$_{orb}$ may be caused by a large 
perturbation of the disk, such as a global density wave 
periodically obscuring our view of the inner regions (\cite{boy01}). \\

The NS involved in the other HMXBs is sometimes a pulsar. In the last 
decade, detailed timing studies were made possible, especially thanks 
to the RXTE satellite. In PSR\,B0540--69 (50ms), the pulse profile,
found to be similar in optical and in different X-ray bands, appears
composed of two gaussians whose ratio does not vary with the energy 
considered (\cite{dep03}). Over the 8 years of RXTE observations,
only one change in the pulsar period (`glitch') was observed (\cite{liv05}). 
Compared to the Crab, this pulsar has a much lower `glitching' activity though otherwise both objects share similar properties (age, magnetic 
field strength, rotation,...). This probably indicates that another physical 
parameter, overlooked up to now, plays here a crucial role. In contrast,
the much older PSR\,J0537--6910 (16.1ms, in N157B) displays a large 
`glitching' activity: 23 glitches were observed in the 7 years of RXTE data 
(\cite{mid06}). The timing analysis yielded interesting results: 
(1) the amplitude $\Delta\nu$ of a glitch is proportional to the interval to 
the next glitch; (2) the longer the time before a glitch, the larger
the change $| \Delta \dot \nu|$ but there is a maximum value for this 
variation; (3) the gain in $| \dot \nu|$ across one glitch is 
not completely given back before the next glitch; (4) microglitches 
often precede large glitches. The overall activity of PSR\,J0537--6910
is much higher than that of the Crab; on the other hand, its glitches are 
smaller but more frequent than for Vela, to which it is often compared. 
The analysis of PSR\,J0537--6910 indicates that glitch models relying on 
sudden onsets are not compatible with the observed glitches and that 
observations of PSR\,J0537--6910 appear in agreement with the picture 
of cracks at the NS surface combined to unstable vortices in the neutron 
superfluid (\cite{mid06}). \\

The HMXB LMC\,X-4 also harbours a pulsar (period
13.5s) in a 1.4d orbit around an O8III star (\cite{neg02}).
The X-ray flux is modulated with a 30d timescale (\cite{lan81}, 
see also \cite{nai03} and \cite{got06}), which likely reflects 
the precession period of the tilted accretion disk as in Her X-1. 
While the orbital period appears very stable, the disk precessing period 
may vary non-uniformly (\cite{tsy05}). 
Pulses and eclipses are also seen in the hard X-ray range by {\sc Integral} 
(\cite{got06}). The eclipses suggest a size for the hard
X-ray emitting region of 0.38\,R$_{\odot}$, i.e. larger than the size of 
a NS and more typical of that of a hot corona. LMC\,X-4 is also varying
at these high energies by an order of magnitude (\cite{got06},
\cite{tsy05}).
Finally, LMC\,X-4 experiences X-ray flares and at these times, mHz 
quasi-periodic oscillations have been detected (\cite{moo01}):
strong, burstlike-features with a timescale of 700--1500s and weak
oscillation with periods of 50--500s. The former could be explained
by beating frequencies between the pulsar frequency and the orbital frequencies
of big clumps on the verge of being accreted, while the latter is more
compatible with Keplerian periods of clumps outside the corotation radius
(\cite{moo01}).

\subsection{LMXBs}
Low-mass X-ray binaries are systems composed of a compact object 
(NS, BH) and a faint, low-mass (generally $<$1\,M$_{\odot}$) companion. 
The X-ray emission is a consequence of the mass transfer via Roche-lobe 
overflow towards the compact object. Only one such object (or maybe two: LMC\,X-2 and possibly RX\,J0532.7$-$6926, \cite{liu07})\footnote{The status 
of RX\,J0532.7$-$6926 is still debated. On the one 
hand, \cite{hab99b} strongly advocate in favor of a LMXB nature on the 
basis of the shape of the X-ray lightcurve. On the other hand, 
\cite{hab99a} and \cite{sas00} only categorize it as a candidate 
LMXB ("LMXB?") while \cite{kah02} mentions the source as a possible 
LMXB or AGN. At the present time, no counterpart has been detected for 
this source. } is known in the LMC, implying that the proportion of LMXBs 
with respect to HMXBs is much smaller in the LMC than in the Galaxy. This 
can be explained by the different star formation history of the two 
galaxies since LMXBs are associated with an older stellar population than 
HMXBs (\cite[Liu \etal\ 2005]{liu05}). \\

LMC\,X-2 is a binary of period 8h (\cite{cor07}), similar in many respects to 
Sco X-1. A monitoring has indicated that, during the bright X-ray states, the
optical lightcurve lags $\lesssim$20s behind the X-ray lightcurve (\cite{mcg03}). 
There thus seems to be some light reprocessing in the system, but 
the location where it takes place is unclear since the lower limit for 
the delay is larger than the light traveltime across the accretion disk
and smaller than the light traveltime to the secondary (\cite{mcg03}).
The X-ray spectrum is well fitted by the combination of a disk 
multi-temperature blackbody and a hot blackbody (kT of 1.5\,keV,
probably associated with regions at or close to the surface of the NS, 
\cite{lav08}).
The inner parts of the disk are not seen: as the source's luminosity is 
close to the Eddington limit, some material can be ejected, which would 
obscure our view towards those inner regions. The Fe\,K$\alpha$ line 
stays undetected while the O\,{\sc viii}\,Ly$\alpha$ is present: this 
is most probably a metallicity effect, which needs to be further 
investigated (\cite{lav08}). 

\section{Massive stars and clusters}

Massive stars, of spectral types O and early B, are soft and moderate
X-ray emitters. In our Galaxy, their overall luminosity scales with their 
bolometric luminosity ($L_X\sim10^{-7}L_{BOL}$) and their spectra reveal 
emission lines from an optically-thin hot plasma with kT=0.3--0.7\,keV.
Such objects are the progenitors of SNe, GRBs, NSs and BHs, and are often
responsible for the presence of diffuse X-ray emissions (SNRs and 
wind-blown bubbles, see Chu, these proceedings). One of their 
main characteristics is the presence of strong stellar winds, driven
through resonant scattering of their intense UV radiation by metals. 
The mass-loss rate and wind velocities of massive stars are typically 
10$^{-6}$\,M$_{\odot}$\,yr$^{-1}$ and 2000\,km\,s$^{-1}$,
respectively. The X-ray emission is generally believed to arise in 
these winds, through collisions of structures travelling at different 
velocities (for a review, see G\"udel \& Naz\'e, in prep.). \\

Since winds are heavily dependent on the metallicity of their 
host galaxy, LMC observations of massive stars are crucially important.
A first test can be performed on Wolf-Rayet stars, the evolved descendents
of O-type stars which display the strongest and densest winds.
WRs mostly come in two flavours, WN if their spectrum is enriched in 
nitrogen and WC in the case of a carbon enrichment. In the Galaxy, no 
single WC star has ever been observed in the X-ray domain, most probably
because of the high absorption of their stellar winds; the situation
for WN is less clear and slight differences in wind structures and
composition could play an important role (for a review, see G\"udel 
\& Naz\'e, in prep.). For the LMC, Guerrero \& Chu (2008a,b) and \cite{gue08c} have 
analyzed all ROSAT, \ch, and \xmm\ observations available, which cover
more than 90\% of the known WRs in the LMC. Of the 125 observed objects, 
only 32 were detected, mostly binaries: the detection 
rate is 50\% for binaries but only 10\% for supposedly single objects.
There are similarities with the Galactic case (non-detection of single WC stars,
binaries preferentially detected) but there are also clear differences.
Notably, the X-ray luminosity and $L_X/L_{BOL}$ ratio are larger for LMC
objects, which could be explained by a lower opacity of the winds. \\

Peculiar phenomena can enhance and harden the stellar X-ray emission: 
(1) in single objects, intense magnetic fields channel the wind streams 
towards the equatorial regions where they collide, producing a very hot 
plasma; (2) in binaries composed of two massive objects, the wind of 
one star can interact with that of the other, again leading to the formation
of a hot plasma (for a review, see G\"udel \& Naz\'e, in prep.). Bright, hard 
X-ray sources associated with massive stars have therefore often been 
attributed to one or the other phenomenon, depending on the authors involved 
(see e.g. for colliding wind binaries in the LMC: \cite{por02}). 
However, one must be cautious about such conclusions. First 
of all, a spectral monitoring in the IR, optical, or UV domain is needed to 
ascertain the binary nature of the object. This is however no definite 
proof of colliding-wind binaries, since magnetic objects (single or 
in binaries) are also overluminous, and it should be noted that not all 
massive binaries are overluminous even if both components
possess significant stellar winds. Second, a monitoring is requested 
in the X-ray domain. Indeed, phase-locked variations are the signature
of peculiar phenomena and help reject the simple line-of-sight coincidence.
The X-ray emission of a magnetic oblique rotator is modulated by the 
(usually short) rotation period of the star, while the X-ray emission 
from wind-wind interactions changes with the orbital period because of varying 
absorptions crossing the line-of-sight or the varying distance between
the two stars (hence a changing strength of the wind-wind collision).
Up to now, variability as just described could be established only in one 
massive system of the MCs, HD\,5980 in the SMC where an \xmm\ monitoring was
performed to ascertain the colliding-wind nature of the emission 
(\cite{naz07}). \\

Massive stars generally reside in clusters. In the LMC, only two of them were 
studied in the X-ray domain: 30 Doradus and N11. A 20ks \ch\ observation of 30 Doradus revealed
180 sources with $L_X>10^{33}$\,erg\,s$^{-1}$, 109 of them being within
30'' of R136 (Townsley \etal\ 2006a,b). Half of the X-ray sources possess 
counterparts at other wavelengths, generally massive stars: some bright, 
hard sources are considered as potential colliding-wind binaries. 
Some non-detection should also be underlined: no star from the embedded 
new stellar generation has been detected and not all early-type objects (e.g. O3)
are detected. A longer exposure (100ks) has now been obtained and is still 
under analysis.
In N11, the coarse spatial resolution of \xmm\ data failed to provide clear 
detections of individual stars, though hints in this direction were found 
in the clusters LH10 and LH13 (\cite{naz04}). A 283ks \ch\ observation 
found 165 point sources in the central area of N11 (clusters LH9/LH10) 
with $L_X>10^{32}$\,erg\,s$^{-1}$ (Chu, Wang, Naz\'e \etal, in prep.). 
Fifteen of these are associated with massive objects (10 with O/WR, 
2 with B stars, 3 with compact groups of massive stars), yielding an 
overall detection rate of 16\% (indeed, the brightest and/or earliest 
objects display the highest detection fraction). Known binaries constitute 
only 20\% of the detected objects: comparing with the clusters's binary 
fraction of about 36\%, this suggests that massive binaries are NOT 
preferentially detected, contrary to what happens in the Galaxy. Moreover,
if the 15 detected objects can be considered as truly typical, the 
$L_X-L_{BOL}$ relation would be 0.4 dex higher in the LMC than in the 
Galaxy, again contrary to expectations. However, it remains to be confirmed
that no peculiar object (magnetic star, colliding wind binary) contaminates
the sample. This is indeed a plausible hypothesis since stars of apparently 
similar spectral types display very different X-ray fluxes (as for 30 Doradus).

  \begin{figure}
  \begin{center}
  \caption{\ch\ observation of N11. Note the large number of point sources, scattered all over the field-of-view, without any correlation with the positions of the clusters (LH10 is in the middle-left of the image and LH9 just below), suggesting a large contamination from background/foreground objects (from Chu, Wang, Naz\'e \etal\ in prep.). }
  \label{N11}
  \end{center}
  \end{figure}

\section{Perspectives}

Many stellar objects emit X-rays. The brightest ones, involving compact
objects (XRBs and SSSs, $L_X\sim10^{36-38}$\,erg\,s$^{-1}$), have been 
detected in the LMC more than 3 decades ago; the current instruments 
have now provided the first detailed timing sequences and high-resolution 
spectra, which often led to changes or refinements of the initial models. 
For these sources, it is now necessary to reach even higher spectral 
resolutions and sensitivity to get more precise observational constraints. 
Moderate X-ray sources such as massive stars (10$^{31-34}$\,erg\,s$^{-1}$)
now enter the picture. At least the brightest examples have been 
detected, notably in 30 Doradus and N11. Beyond enlarging the number of
X-ray detections, the future lies in getting high-resolution spectroscopy 
of LMC massive stars: since their X-ray emission is linked to their
stellar winds, which crucially depend on metallicity,  high-res data
are necessary to test our X-ray generation models. Indeed, in the 
Galaxy, such high-res observations have already initiated a shift in thought 
- but it is essential to check the new theories in a different metallicity 
environment like that of the LMC.
Finally, the future generation of X-ray telescopes should be able to 
detect even fainter X-ray sources such as low-mass/coronal sources 
(10$^{26-33}$\,erg\,s$^{-1}$) and young pre-main sequence objects 
(flaring T Tauri stars can reach 10$^{31-32}$\,erg\,s$^{-1}$)\footnote{Note that cataclysmic variables (CVs) can also be rather bright X-ray sources, but they are difficult to identify in the Magellanic Clouds due to the faintness of their optical counterparts. Currently, there are no clear identification of a LMC X-ray source as a CV and this is why CVs were not considered in this review.}. \\

Once all this is accomplished, a full picture of the LMC at high energies 
will be available. Of course, acquiring such data is not a simple question of 
`filling the catalogues'. It must always be kept in mind that, with its 
lower metallicity, the LMC provides a crucial test of theoretical models 
- see e.g. the case of massive stars. The astronomical community should
thus promote the advent of a new generation of X-ray facilities possessing 
three concommitant characteristics: high sensitivity (to detect the faintest 
X-ray sources in nearby galaxies), high spatial resolution (to disentangle 
blended stellar objects in nearby galaxies), and high spectral resolution.\\

{\it Acknowledgments}\\
YN acknowledges financial support from the Fonds de la Recherche Scientifique (FRS-FNRS Belgium), the University of Li\`ege (through the `patrimoine-ULg' grants), the organizers of the Symposium, and the PRODEX XMM and Integral contracts. \\

\end{document}